\documentclass[usenatbib,useAMS]{mn2e}
\usepackage{natbib}
\usepackage{amsmath}
\usepackage{amssymb}
\usepackage{graphicx}
\usepackage{subfigure}
\usepackage{caption}
\usepackage{color}
\usepackage{epsfig}
\allowdisplaybreaks[1]

\newcommand{\beq}{\begin{equation}}
\newcommand{\eeq}{\end{equation}}

\newcommand{\apj}{ApJ}
\newcommand{\apjl}{ApJL}

\newcommand{\araa}{ARA$\&$A}

\newcommand{\mnras}{MNRAS}
\newcommand{\pasp}{PASP}

\newcommand{\nat}{Nature}

\title[Fast Luminous Blue Transients from Newborn BHs]
{Fast Luminous Blue Transients from Newborn Black Holes}
\author[K. Kashiyama,  E. Quataert]{Kazumi Kashiyama$^{1,2}$\thanks{E-mail:kashiyama@berkeley.edu} and Eliot Quataert$^{1}$\\
$^{1}$Department of Astronomy; Theoretical Astrophysics Center; University of California, Berkeley, Berkeley, CA 94720, USA\\
$^{2}$Einstein fellow}

\begin{document}

\date{Accepted 2015 ???. Received 2015 ???; in original form 2015 ???}

\pagerange{\pageref{firstpage}--\pageref{lastpage}} \pubyear{2015}

\maketitle

\label{firstpage}

\begin{abstract}
Newborn black holes in collapsing massive stars can be accompanied by a fallback disk. 
The accretion rate is typically super-Eddington and strong disk outflows are expected. 
Such outflows could be directly observed in some failed explosions of compact (blue supergiants or Wolf-Rayet stars) progenitors, 
and may be more common than long-duration gamma-ray bursts. 
Using an analytical model, we show that the fallback disk outflows produce blue UV-optical transients 
with a peak bolometric luminosity of $\sim 10^{42-43} \ \rm erg \ s^{-1}$ (peak R-band absolute AB magnitudes of $-16$ to $-18$) 
and an emission duration of $\sim$ a few to $\sim$ 10 days. 
The spectra are likely dominated intermediate mass elements, but will lack much radioactive nuclei and iron-group elements. 
The above properties are broadly consistent with some of the rapid blue transients detected by Pan-STARRS and PTF. 
This scenario can be distinguished from alternative models using radio observations within a few years after the optical peak. 
\end{abstract}


\section{introduction}
Stellar-mass black holes~(BHs) are now ubiquitously found in X-ray binaries.
Such BHs are formed in collapsing massive stars, but many questions remain about how the progenitor properties connect to those of the resulting BHs:  
Which progenitors produce BHs not neutron stars (NS)~\citep{Clausen_et_al_2014}?
What is the initial mass function of BHs~\citep{Kochanek_2015}?
What is the initial spin distribution?
Modern surveys can directly address these questions by catching BH formation in the local universe. 
Roughly a million supergiants within $\sim 10 \ \rm Mpc$ are being monitored with a cadence of a few months~\citep{Kochanek_et_al_2008,Gerke_et_al_2014}. 
One can expect $\sim 1 \ \rm yr^{-1}$ collapse among this sample.  

Some fraction of collapsing massive stars are expected to lead to BH formation with a failed supernova (SN) explosion (in the sense of no traditional $\gtrsim 10^{50} \ \rm erg$ roughly spherical explosion).   
Even such events may, however, be accompanied by an electromagnetic counterpart~\citep{Kochanek_et_al_2008}.
For example, the gravitational mass loss of $\sim 0.1 \ M_\odot$ via neutrinos in the proto-NS phase 
can drive a weak explosion of $\sim 10^{47} \ \rm erg$ in the case of a red supergiant~(RSG).  
This may result in a luminous red nova \citep{Nadezhin_1980,Lovegrove_Woosley_2013} with shock breakout emission~\citep{Piro_2013}.

\begin{figure*}
\includegraphics[width=2.0\columnwidth]{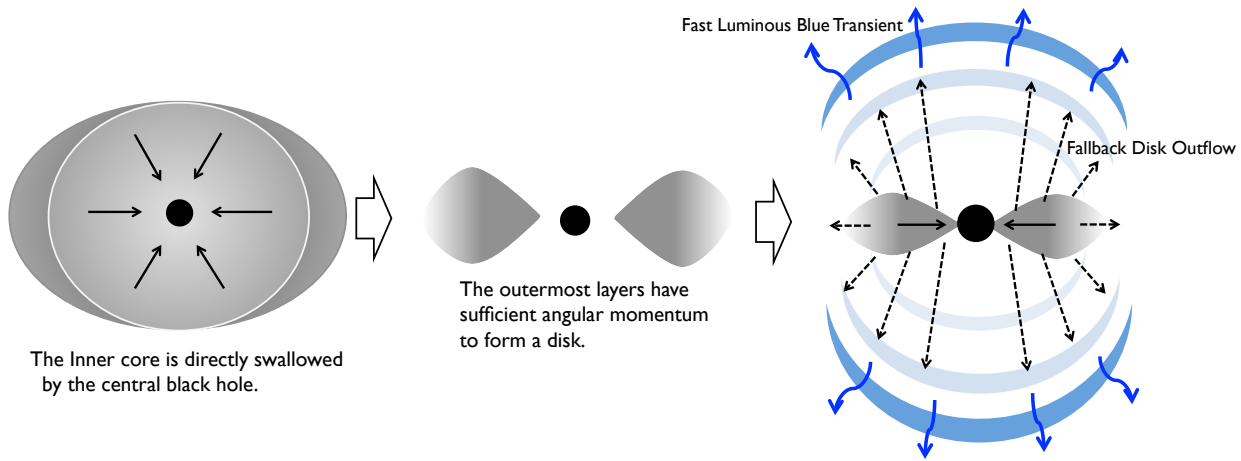}
\caption{
Schematic picture of failed supernova model for fast luminous blue transients.
}
\label{fig:sche}
\end{figure*}

In this paper, we first review the possible diversity of BH formation and its electromagnetic counterpart~(Sec. \ref{sec:bhf_div}). 
Then, we focus on a relatively unexplored, but potentially promising observational target;  BH fallback disk outflow. 
We consider the limit in which the inner core of the progenitor is directly swallowed by the BH, but the outermost layers have sufficient angular momentum to form a disk~\citep[``type III collapsars" in][]{Woosley_Heger_2012} 
(see Fig. \ref{fig:sche}). 
The accretion rate is still super-Eddington, and a strong outflow is expected to be launched from the disk~\citep{Ohsuga_et_al_2005,Sadowski_et_al_2014,Jiang_et_al_2014}. 
If there is no significant quasi-spherical explosion initially, this outflow can be directly seen by distant observers. 
This situation can be realized in the collapse of e.g., Wolf-Rayet stars~(WRs) and blue supergiants (BSGs), 
and may in fact be rather common~\citep{Woosley_Heger_2012,Perna_et_al_2014}. 
We analytically calculate the emission from fallback disk outflows, 
and show that they can be observed as blue UV-optical transients with a peak bolometric luminosity of $\sim 10^{42-43} \ \rm erg \ s^{-1}$ 
and an emission duration of $\sim$ a few to $\sim$ 10 days~(Sec. \ref{sec:lbn}). 
Such emissions may explain some of the rapid transients recently discovered by Panoramic Survey Telescope \& Rapid Response System~(Pan-STARRS) and Palomar Transient Factory~(PTF)~(Sec. \ref{sec:dis}).  
Our scenario can be tested by detecting a radio afterglow within a few years after the optical peak. 

\section{Diversity of Black Hole Formation}\label{sec:bhf_div}
Stellar-mass BHs are predominately formed in the core collapse of massive stars. 
For stars with zero-age-main-sequence~(ZAMS) masses of $\gtrsim 10 M_\odot$, the iron core collapses once its mass exceeds the Chandrasekhar limit, forming a proto-NS. 
The proto-NS cools via intense neutrino emission of $\sim 10^{53} \ \rm erg$, which is believed to ultimately power the SN explosion, at least in some progenitors
\citep[e.g.,][]{OConnor_Ott_2011,Ugliano_et_al_2012,Horiuchi_et_al_2014,Pejcha_Thompson_2015}.  
If, however, the accretion shock onto the proto-NS stalls and never reverses to unbind the stellar envelope, the continued accretion eventually leads the proto-NS to collapse into a BH. 
We are interested in the latter case in this paper.  
Even in this case, however, if the progenitor is a RSG, the change in core mass associated with neutrino radiation leads to a weak shock 
propagating out through the stellar envelope that unbinds $\sim$ a few $M_\odot$ of the envelope with a relatively small velocity of $\sim 100 \ \rm km \ s^{-1}$~\citep{Nadezhin_1980,Lovegrove_Woosley_2013}.
The weak shock also heats up the ejecta, leading to a slow red transient lasting for $\sim 100 \ \rm days$ with a bolometric luminosity of $\sim 10^{40} \ \rm erg \ s^{-1}$.

WR or BSG progenitors have steeper density gradients and more tightly bound envelopes than RSGs. 
As a result, it is likely that there is much less mass lost in quasi-spherical ejecta as a response to the neutrino radiation of the proto-NS binding energy (though this remains to be demonstrated by detailed calculations).  
If there is BH formation associated with the failed explosion of WR or BSG progenitors, such events may thus have little electromagnetic signature in the absence of significant rotation. 
In the presence of significant angular momentum, however, the accretion disk that forms during collapse can power electromagnetic emission that would accompany BH formation in nominally failed explosions.

WR or BSG progenitors with significant angular momentum are also the leading progenitors for long-duration (and ultra-long duration) gamma-ray bursts~(GRBs)~\citep[e.g.,][]{MacFadyen_Woosley_1999,Kashiyama_et_al_2013}.  
These are observationally associated with robust, energetic explosions~\citep[e.g.,][]{Woosley_Bloom_2006}.  
It is not guaranteed, however, that every, or even the majority of, WR or BSG progenitors with significant angular momentum produce successful GRBs. 
For example, the powerful jets that produce GRBs may require large-scale magnetic flux in the stellar progenitor as well as rapid rotation~\citep{Tchekhovskoy_Gianios_2014}. 
Since large-scale magnetic flux tends to slow down the rotation of the core during stellar evolution, it could be that the combination of conditions required to produce luminous GRBs is somewhat rare.  
In this paper, we argue that an alternative electromagnetic counterpart associated with BH formation during the collapse of rapidly rotating WR or BSG progenitors is a fast luminous UV-optical transient 
broadly similar to some events discovered by Pan-STARRs and PTF in the last $\sim 5$ years~\citep{Drout_et_al_2014}.

\section{Fast Luminous Blue Transients}\label{sec:lbn}
Hereafter, we consider the electromagnetic counterpart of collapse 
in which the bulk of the progenitor directly accretes onto the BH, and the outer most layers can form a fallback disk (see Fig. \ref{fig:sche}).
We focus on the fallback disk outflow and the cooling emission in the course of its expansion, which can be seen directly by observers.\footnote{
Similar emission has been discussed in the context of fallback accretion disk in tidal disruption of stars by super-massive BHs~\citep{Strubbe_Quataert_2009,Strubbe_Quataert_2011} 
and of NSs in NS-BH binaries~\citep{Rossi_Begelman_2009}.}
This situation can be realized e.g., for WRs in close binaries or BSGs with little mass loss during stellar evolution~\citep{Woosley_Heger_2012}. 
Note that some RSGs may also lead to``failed" explosions. 
However, in this case, spherical mass ejection of $\sim$ a few $M_\odot$ driven by the neutrino mass loss is more probable, 
and the disk outflows predicted here are likely hidden by this ejecta. 
Any electromagnetic source associated with the fallback disk would be powered by the kinetic energy of the outflow thermalized via interaction with this ejecta~\citep{Dexter_Kasen_2013}. 

Once the SN shock stalls, the outer layers of the progenitor fall back to the central BH. 
If the outer layers have sufficient angular momentum, they form an equatorial torus at the circularization radius, 
\begin{equation}\label{eq:r_0}
r_0 \approx f_r \times \frac{2GM_{\rm BH}}{c^2} \sim 3 \times 10^{7} \ {\rm cm} \ \left(\frac{f_r}{10}\right) \left(\frac{M_{\rm BH}}{10 \ M_\odot}\right). 
\end{equation} 
We focus on marginal cases in which the circularization radius is not much larger than the innermost circular orbit of the BH ($f_r \sim 10-100$); 
we show below (Fig. \ref{fig:Rab_fr}) that larger circularization radii likely lead to fainter more slowly evolving transients.
The fallback rate can be estimated as 
$\dot{M}_{\rm d} \approx M_{\rm d}/t_{\rm acc}$, or 
\begin{eqnarray}\label{eq:M_acc}
\dot{M}_{\rm d} &\sim& 3 \times 10^{-5} \ {\rm M_\odot \ s^{-1}} \notag \\ 
&& \times \left(\frac{M_{\rm d}}{1 \ M_\odot}\right) \left(\frac{R_*}{10^{12} \ \rm cm}\right)^{-3/2} \left(\frac{M_{\rm BH}}{10 \ M_\odot}\right)^{1/2},
\end{eqnarray}
where $t_{\rm acc} \approx \pi(R_*{}^3/8 G M_{\rm BH})^{1/2}$, or 
\begin{equation}\label{eq:t_acc}
t_{\rm acc} \sim 3 \times 10^{4} \ {\rm s} \ \left(\frac{R_*}{10^{12} \ \rm cm}\right)^{3/2} \left(\frac{M_{\rm BH}}{10 \ M_\odot}\right)^{-1/2}
\end{equation}
is the free fall timescale, $R_*$ is the radius of the outermost layer, and $M_{\rm d}$ is the total mass of the disk. 
The torus is optically and geometrically thick, trapping the heat generated by the fallback material. 

The disk accretes once the angular momentum is effectively transported by e.g., magnetorotational instability~(MRI).  
\cite{Proga_Begelman_2003a,Proga_Begelman_2003b} simulated accretion of low angular momentum gas in a scenario qualitatively analogous to that considered here.  
They showed that the MRI redistributes angular momentum during the circularization, leading to dissipation which powers both accretion and an outflow.   
In our scenario, the viscous time of the disk is much shorter than the fallback time scale. 
Thus, the accretion rate is essentially given by the fallback rate (Eq. \ref{eq:M_acc}), 
which is typically larger than the Eddington accretion rate, 
$\dot{M}_{\rm Edd} = 4 \pi G M_{\rm BH}/c \kappa \sim 1 \times 10^{-15} \ M_\odot \ {\rm s^{-1}} \ (\kappa/0.2 \ {\rm cm^2 \ g^{-1}})^{-1} (M_{\rm BH}/10 M_\odot)$.
Note that the opacity $\kappa \sim 0.1, \ 0.2$ and $0.4 \ \rm cm^{2} \ g^{-1}$ corresponds to electron scattering for singly ionized helium, fully ionized helium and hydrogen, respectively.
The accretion rate is also below the accretion rate at which there is significant neutrino cooling~\citep{Chen_Beloborodov_2007}.
In this case, one can expect a strong radiation-driven outflow from the disk.
Such outflows have been also confirmed by numerical simulations~\citep{Ohsuga_et_al_2005,Sadowski_et_al_2014,Jiang_et_al_2014}.

We model the fallback disk outflow as follows. 
First, a fraction $f_{\dot{M}} < 1$ of the accreting mass is loaded on the outflow, $\dot{M}_{\rm out} = f_{\dot{M}} \times \dot{M}_{\rm d}$ or, 
\begin{eqnarray}
\dot{M}_{\rm out} &\sim& 3 \times 10^{-6} \ {\rm M_\odot \ s^{-1}} \ \left(\frac{f_{\dot{M}}}{0.1} \right)  \\ \notag
&& \times  \left(\frac{M_{\rm d}}{1 \ M_\odot}\right)  \left(\frac{R_*}{10^{12} \ \rm cm}\right)^{-3/2} \left(\frac{M_{\rm BH}}{10 \ M_\odot}\right)^{1/2}. 
\end{eqnarray}
Second, the outflow velocity is approximately the escape velocity, $\bar v_{\rm out} \approx (2G M_{\rm BH}/r_0)^{1/2}$, or 
\begin{equation}\label{eq:var_v_out}
\bar v_{\rm out} \sim 1 \times 10^{10} \ {\rm cm \ s^{-1}}  \left(\frac{f_r}{10}\right)^{-1/2}. 
\end{equation}
Finally, we assume that the outflow is isotropic, although in reality it will be moderately bipolar. 

Next, let us describe the density and temperature profile in the outflow, which are crucial for quantifying the electromagnetic emission.  
After the launch, the outflow expands into the surrounding medium.
For $t \lesssim t_{\rm acc}$, the accretion rate is almost constant, and the outflow is approximately a steady wind. 
The density structure can be described as $\rho \approx \rho_0 (r/r_0)^{-2}$, 
where $\rho_0 \approx \dot{M}_{\rm out}/4\pi r_0^2 \bar v_{\rm out}$, or 
\begin{eqnarray}
\rho_0 &\sim& 60 \ {\rm g \ cm^{-3}} \ \left(\frac{f_r}{10}\right)^{-3/2}  \left(\frac{f_{\dot{M}}}{0.1}\right)  \notag \\
&& \times \left(\frac{M_{\rm d}}{1 \ M_\odot}\right)  \left(\frac{R_*}{10^{12} \ \rm cm}\right)^{-3/2} \left(\frac{M_{\rm BH}}{10 \ M_\odot}\right)^{-3/2}, 
\end{eqnarray}
is the density of the outflow at $r = r_0$. 
Since the outflow is initially highly optically thick, the temperature evolves adiabatically, $T \propto \rho^{1/3} \propto r^{-2/3}$, thus $T \approx T_0 (r/r_0)^{-2/3}$,
where $T_0 \approx (\dot{M}_{\rm out} v_{\rm out}/8 \pi a r_0^2)^{1/4}$, or 
\begin{eqnarray}\label{eq:T_0}
T_0 &\sim& 8 \times 10^8 \ {\rm K} \ \left(\frac{f_r}{10}\right)^{-5/8} \left(\frac{f_{\dot M}}{0.1}\right)^{1/4}  \notag \\
&& \times \left(\frac{M_{\rm d}}{1 \ M_\odot}\right)^{1/4}   \left(\frac{R_*}{10^{12} \ \rm cm}\right)^{-3/8}  \left(\frac{M_{\rm BH}}{10 \ M_\odot}\right)^{-3/8}. 
\end{eqnarray}
We note that the gas temperature in the disk is $T_{\rm d} \approx f_{\dot M}^{-1/4} T_0 \lesssim$ a few $10^9 \ \rm K$. 
In this case, heavy nuclei up to at most O, Ne, and Mg can be synthesized inside the disk, but not Fe group elements. 
Hence, there is no radioactivity in the outflow. 
The above nuclear burning only occurs in the inner most disk, where the enthalpy is likely larger than the nuclear energy released,  
so that the nuclear reactions are not dynamically important~\citep{Fermandez_Metzger_2013}.

At $t \gtrsim t_{\rm acc}$, the accretion rate decreases significantly,  
and the outflow essentially decouples from the disk. 
Then, the outflow ejecta will expand in a homologous manner, $r/t \approx v$. 
The density profile of the homologous ejecta can be described as
\begin{equation}\label{eq:rho_homo}
\rho \approx \rho'_0 \left( \frac{t}{t_{\rm acc}}\right)^{-3} \left( \frac{v}{v_{\rm out, min}}\right)^{-\xi}. 
\end{equation}
We determine the normalization of the density by mass conservation, 
$\int^{r_{\rm max}}_{r_{\rm min}} 4 \pi r^2 \rho dr \approx f_{\dot M} M_{\rm d}$, which yields 
\begin{eqnarray}
\rho'_0 &\sim& 4 \times 10^{-12} \ {\rm g \ cm^{-3}} \ \left(\frac{f_r}{10}\right)^{3/2}  \left(\frac{f_{\dot{M}}}{0.1}\right) \notag \\
&& \times \left(\frac{M_{\rm d}}{1 \ M_\odot}\right)  \left(\frac{R_*}{10^{12} \ \rm cm}\right)^{-9/2} \left(\frac{M_{\rm BH}}{10 \ M_\odot}\right)^{3/2}.
\end{eqnarray}
Here, $r_{\rm max} \approx v_{\rm out, max} t_{\rm acc}$, $r_{\rm min} \approx v_{\rm out, min} t_{\rm acc}$, 
$v_{\rm out, max} = f_{v, \rm max} \bar v_{\rm out}$, and $v_{\rm out, min} = f_{v, \rm min} \bar v_{\rm out}$.  
In this paper, we choose $f_{\rm v, max} \gtrsim 1$, $f_{\rm v, min} \lesssim 1$, and $\xi > 2$ so as to satisfy the energy conservation 
i.e., $\int^{r_{\rm max}}_{r_{\rm min}} (4 \pi r^2 \times \rho v^2/2) dr \approx f_{\rm \dot M} M_{\rm d} \bar v_{\rm out}{}^2/2$.  
We note that the internal energy of the shell is subdominant at $r \approx r_{\rm min}$ due to adiabatic cooling.
To obtain $f_{\rm out, max}$, $f_{\rm out, min}$, and $\xi$ consistently, one has to perform numerical simulations, 
but the basic characteristics of the optical emission are not so sensitive to these parameters. 
We take $f_{v, \rm min} = 0.7$, $f_{v, \rm max} = 1.4$, and $\xi = 3.75$ as fiducial choices. 
As long as the ejecta is almost adiabatic, the temperature profile can be described as 
\begin{equation}
T \approx  T'_0 \left( \frac{t}{t_{\rm acc}}\right)^{-1} \left( \frac{v}{v_{\rm out, min}}\right)^{-\xi/3}.
\end{equation}
where $T'_0 \approx T_0(\rho'_0/\rho_0)^{1/3}$, or
\begin{eqnarray}
T'_0 &\sim& 3 \times 10^4 \ {\rm K} \ \left(\frac{f_r}{10}\right)^{3/8}  \left(\frac{f_{\dot{M}}}{0.1}\right)^{1/4} \notag \\
&& \times  \left(\frac{M_{\rm d}}{1 \ M_\odot}\right)^{1/4}  \left(\frac{R_*}{10^{12} \ \rm cm}\right)^{-11/8} \left(\frac{M_{\rm BH}}{10 \ M_\odot}\right)^{5/8}.
\end{eqnarray}

We now estimate the electromagnetic emission from the outflow. 
As the outflow expands, the photons diffusively come out from the diffusion radius, $r_{\rm dif}$, 
which is defined by the radius where the diffusion time of the photon is comparable to the expansion time of the outflow, i.e., $t = t_{\rm dif}$, where 
\begin{equation}\label{eq:emi1}
t_{\rm dif} = \tau \frac{\Delta r}{c},
\end{equation}
\begin{equation}\label{eq:emi2}
\tau = \int^{r_{\rm max}}_{r_{\rm dif}} \kappa \rho dr,
\end{equation}
is the optical depth, and 
\begin{equation}\label{eq;emi3}
\Delta r = r_{\rm max} - r_{\rm dif}
\end{equation}
is the diffusion width.  
For each $t$, one can calculate $r_{\rm dif}$ and $\Delta r$ from Eqs. (\ref{eq:emi1}-\ref{eq;emi3})~\citep{Kisaka_et_al_2014}.\footnote{
We set $v_{\rm out, max} = 2 v_{\rm out, min}$. In this case, the thick diffusion phase in the homologous shell discussed in \cite{Kisaka_et_al_2014} does not appear.}  
The emission is approximately thermal with a temperature of $T_{\rm obs}$ at $r = r_{\rm dif}$. 
The bolometric luminosity of the emission is thus given by 
\begin{equation} 
L_{\rm bol} \approx 4 \pi a T_{\rm obs}^4 r_{\rm dif}^2 \frac{\Delta r}{t}.  
\end{equation}

The evolution of the emission as a function of time can be approximately described as follows. 
Just after the outflow is launched, photons only come out from a thin outer layer of the expanding wind profile, 
and the diffusion radius effectively coincides with the outer edge; 
\begin{equation}\label{eq:rph_early}
r_{\rm dif} \approx \bar v_{\rm out}t.
\end{equation}
In this case, from Eqs. (\ref{eq:emi1}-\ref{eq;emi3}), the diffusion width is approximately given by  
\begin{equation} 
\Delta r \approx \sqrt{\frac{ct}{\kappa \rho(r_{\rm dif})}} \propto t^{3/2}.
\end{equation}
Note that $\rho(r_{\rm dif}) \propto r_{\rm dif}^{-2} \propto t^{-2}$ in this phase.   
The temperature and bolometric luminosity evolve as  
\begin{equation}\label{eq:Tph_early}
T_{\rm obs} \approx T_0 \left( \frac{r_{\rm dif}}{r_0}\right)^{-2/3} \propto t^{-2/3}, 
\end{equation} 
and
\begin{equation}\label{eq:Lbol_early}
L_{\rm bol} \propto t^{-1/6},
\end{equation}
respectively. 
The homologous expansion sets in at $t \approx t_{\rm acc}$, which is $\lesssim$ a day for our fiducial parameters (Eq. \ref{eq:t_acc}). 
Then, one has to consider photon diffusion in the density profile of Eq. (\ref{eq:rho_homo}).
In our case, the diffusion radius initially practically coincides with the outer edge at $t \approx t_{\rm acc}$;  
\begin{equation}\label{eq:rph_homo}
r_{\rm dif} \approx v_{\rm out, max} t,
\end{equation}
and the diffusion width can be described as 
\begin{equation}
\Delta r \approx \sqrt{\frac{ct}{\kappa \rho(r_{\rm dif})}} \propto t^2. 
\end{equation}
Now $\rho(r_{\rm dif}) \propto r_{\rm dif}{}^{-3} \propto t^{-3}$. 
Accordingly, the observed temperature and bolometric luminosity evolve as 
\begin{equation}\label{eq:Tph_homo}
T_{\rm obs} \propto t^{-1},
\end{equation}
\begin{equation}\label{eq:Lbol_homo}
L_{\rm bol} \propto t^{-1}. 
\end{equation}
The energy diffuses throughout the entire homologous shell when $r_{\rm dif} \approx r_{\rm min}$, i.e.,
\begin{eqnarray}\label{eq:t_tr}
t_{\rm p} &\approx& \sqrt{\frac{\kappa \rho'_0 t_{\rm acc}^3 v_{\rm out, min}^2}{c}} \times \sqrt{\frac{1-(f_{v, \max}/f_{v, \min})^{1-\xi}}{\xi -1}} \notag \\ 
&\sim& 1.1 \ {\rm days} \ \left(\frac{f_r}{10}\right)^{1/4}  \left(\frac{f_{\dot{M}}}{0.1}\right)^{1/2} \notag \\
&& \times \left(\frac{M_{\rm d}}{1 M_\odot}\right)^{1/2} \left(\frac{\kappa}{0.2 \ {\rm cm^2 \ g^{-1}}}\right)^{1/2}.   
\end{eqnarray}
The observed emission radius, temperature, and bolometric luminosity at $t = t_{\rm p}$ can be estimated as  
\begin{eqnarray}\label{eq:rph_tr}
r_{\rm dif, p} &\approx& v_{\rm out, min} t_{\rm p} \notag \\ 
&\sim& 6 \times 10^{14} \ {\rm cm} \ \left(\frac{f_r}{10}\right)^{-1/4}   \left(\frac{f_{\dot{M}}}{0.1}\right)^{1/2} \notag \\
&& \times \left(\frac{M_{\rm d}}{1 M_\odot}\right)^{1/2} \left(\frac{\kappa}{0.2 \ {\rm cm^2 \ g^{-1}}}\right)^{1/2}, 
\end{eqnarray}
\begin{eqnarray}\label{eq:Tph_tr}
T_{\rm obs, p} &\approx& T'_0 \left(\frac{t_{\rm p}}{t_{\rm acc}}\right)^{-1} \notag \\ 
&\sim& 1 \times 10^{4} \ {\rm K} \  \left(\frac{f_r}{10}\right)^{1/8}  \left(\frac{f_{\dot{M}}}{0.1}\right)^{-1/4} \notag \\ 
&& \times \left(\frac{M_{\rm d}}{1 M_\odot}\right)^{-1/4} \left(\frac{R_*}{10^{12} \ \rm cm} \right)^{1/8}  \left(\frac{M_{\rm BH}}{10 M_\odot}\right)^{1/8}  \notag \\
&& \times \left(\frac{\kappa}{0.2 \ {\rm cm^2 \ g^{-1}}}\right)^{-1/2}, 
\end{eqnarray}
\begin{eqnarray}\label{eq:Lbol_tr}
L_{\rm bol, p} &\approx&  \frac{4 \pi a T_{\rm obs, p}^4 r_{\rm dif, p}^3}{t_{\rm p}} \notag \\ 
&\sim& 2 \times 10^{42} \ {\rm erg \ s^{-1}} \ \left(\frac{f_r}{10}\right)^{-1/2}  \notag \\ 
&& \times \left(\frac{R_*}{10^{12} \ \rm cm} \right)^{1/2}  \left(\frac{M_{\rm BH}}{10 M_\odot}\right)^{1/2} \notag \\
&& \times \left(\frac{\kappa}{0.2 \ {\rm cm^2 \ g^{-1}}}\right)^{-1},
\end{eqnarray}
respectively. 
We note that $t_{\rm p}$ does not depend on $R_*$, and on the other hand, $L_{\rm bol, p}$ does not depend on $M_{\rm d}$. 
In addition, $T_{\rm obs, p}$ depends only weakly on parameters~(Eq. \ref{eq:Tph_tr}). 
Moreover, it is $\sim 10^{4} \ \rm K$ so that optical emission will peak soon after $t_{\rm p}$. 
Eq. (\ref{eq:Lbol_tr}) can be written in an intuitive form,
\begin{equation}\label{eq:Lbol_tr_2}
L_{\rm bol, p} \approx {\cal C} \times E_{\rm int, 0}\left(\frac{\bar v_{\rm out} t_{\rm acc}}{r_{\rm 0}}\right)^{-2/3} \left(\frac{t_{\rm p}}{t_{\rm acc}}\right)^{-1} \frac{1}{t_{\rm p}}
\end{equation}
where $E_{\rm int, 0} = f_{\dot M} M_{\rm d}/\rho_0 \times a T_0^4$ is the initial internal energy of the outflow and 
${\cal C} = (1/f_{\rm v, min}) \times [(3-\xi)/((f_{\rm v, max}/f_{\rm v, min})^{3-\xi}-1)]^{4/3}$ is a factor of order unity determined by the profile of the homologous shell. 
The internal energy decreases with time $\propto t^{-2/3}$ for $t \lesssim t_{\rm acc}$ and $\propto t^{-1}$ for $t_{\rm acc} \lesssim t \lesssim t_{\rm p}$; 
most of the energy is radiated over the timescale $\sim t_{\rm p}$. 

Note that Eq. (\ref{eq:Lbol_tr_2}) is different from the standard order of magnitude expression for the thermal energy radiated by a stellar explosion, which is
\begin{equation}\label{eq:Lbol_tr_3}
L_{\rm bol, p} \approx E_{\rm int, 0} \left(\frac{\bar v_{\rm out} t_{\rm p} }{r_0}\right)^{-1} \frac{1}{t_{\rm p}},
\end{equation}
where $r_0$ should now be interpreted as the radius of the pre-explosion star.   
The difference between Eqs. (\ref{eq:Lbol_tr_2}) and (\ref{eq:Lbol_tr_3}) is that in the scenario explored here, 
the outflow is powered by a wind for $t \lesssim t_{\rm acc}$ and is only homologous for $t \gtrsim t_{\rm acc}$. 
This changes the magnitude of the adiabatic losses.

At $t > t_{\rm p}$, the emission from the shell decreases quickly; 
we assume $T_{\rm obs} \propto t^{-1}$, $L_{\rm bol} \propto t^{-2}$ in the simple numerical models presented below (Figs. \ref{fig:Rab}-\ref{fig:Rab_fr}).
However, our estimates at these times are not very accurate because the ejecta becomes highly non-adiabatic and radiates the majority of its thermal energy.
Also, if He (ONeMg) nuclei dominate the opacity of the ejecta, the opacity significantly decreases once the temperature drops down to $\sim 9000~(6000) \ \rm K$~\citep[e.g.,][]{Kleiser_Kasen_2014,Piro_Morozova_2014}.
As in \cite{Kleiser_Kasen_2014}, this recombination will cause the light curves to fade rapidly, which typically occurs after $t \gtrsim t_{\rm p}$ in our cases. 
Moreover, the outflow becomes transparent i.e., $\tau \sim 1$ at $t \approx (c/v_{\rm min})^{1/2} t_{\rm p} \sim {\rm a \ few} \times t_{\rm p}$, and the thermal emission will rapidly fade thereafter.

\begin{table}
\centering
\captionsetup{justification=centering}
\caption{Model parameters of fallback disk outflow used in Fig. \ref{fig:rph_Tph_Lbol} and \ref{fig:Rab}.}\label{table:para}
\begin{tabular}{l*{2}{c}}
\hline
            &  WR & BSG \\
\hline
Stellar radius ($R_*$) & $10^{11} \ {\rm cm}$ &  $10^{12} \ {\rm cm}$ \\
Disk mass ($M_{\rm d}$)  & $1  M_\odot$ & $5 M_\odot$ \\
Black-hole mass ($M_{\rm BH}$) & $15 M_\odot$ & $40 M_\odot$ \\
Mass loading ($f_{\dot{M}}$) & 0.1 & 0.1 \\
Launching radius ($f_r$) & 10 & 10 \\
Opacity ($\kappa$) & $0.1  \ {\rm cm^2 \ g^{-1}}$ & $0.2 \ {\rm cm^2 \ g^{-1}}$ \\
Minimum velocity ($f_{v, \rm min}$) & 0.7 & 0.7 \\
Maximum velocity ($f_{v, \rm max}$) & 1.4 & 1.4 \\
Density profile ($\xi$) & 3.75 & 3.75 \\
\hline
\end{tabular}
\end{table}

\begin{figure}
\includegraphics[width=1.0\columnwidth]{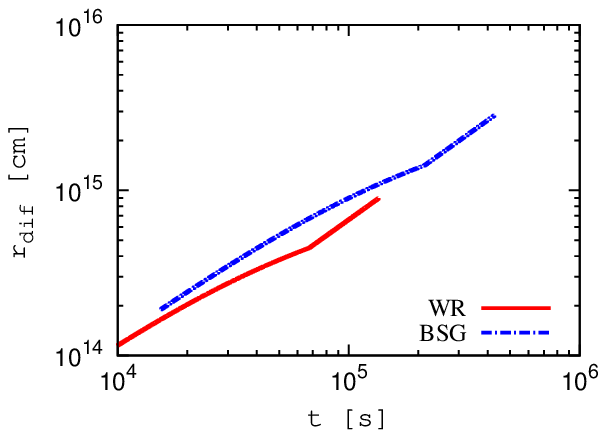}
\includegraphics[width=1.0\columnwidth]{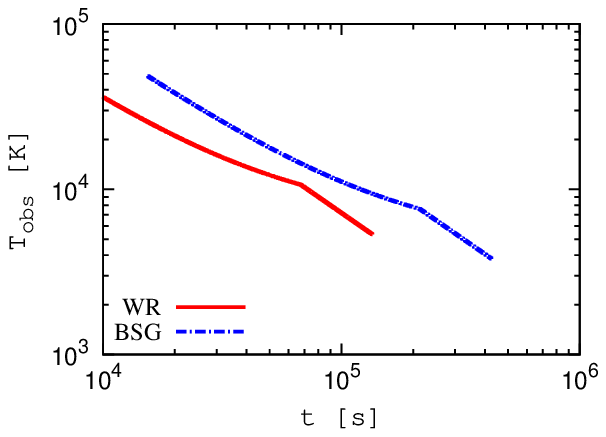}
\includegraphics[width=1.0\columnwidth]{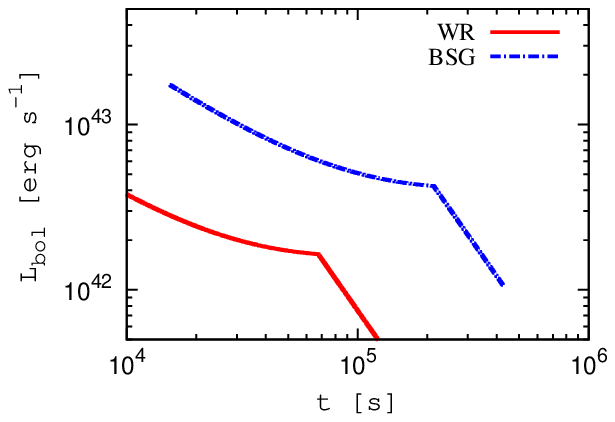}
\caption{
Time evolution of emission radius (top), observed temperature (middle), and bolometric luminosity (bottom) of fallback disk outflow emission.
The solid red and dotted-dash blue curves represents our fiducial models of Wolf-Rayet~(WR) and blue supergiants~(BSG) (Table \ref{table:para}), respectively.  
}
\label{fig:rph_Tph_Lbol}
\end{figure}

\begin{figure}
\includegraphics[width=1.0\columnwidth]{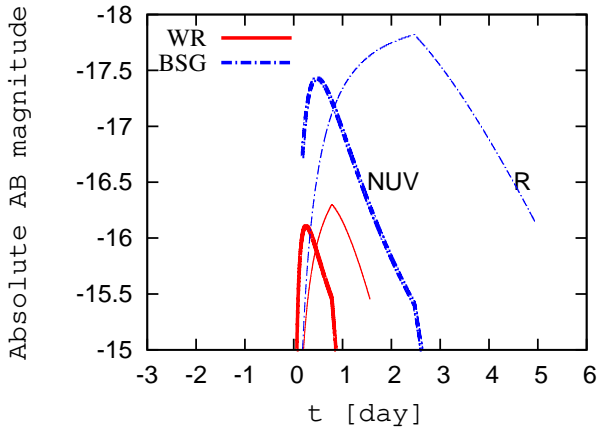}
\caption{
Absolute AB magnitude of fallback disk outflow emission. 
The thick and thin curves show the NUV and R band magnitude, 
and the solid red and dotted-dash blue curves represents our fiducial model of Wolf-Rayet~(WR) and blue supergiants~(BSG) (Table \ref{table:para}), respectively.  
Note that $t = 0$ corresponds to the initial launching of the outflow.
}
\label{fig:Rab}
\end{figure}

\begin{figure}
\includegraphics[width=1.0\columnwidth]{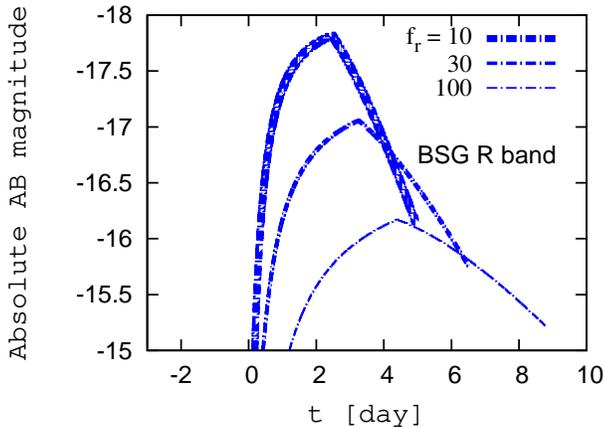}
\caption{
Absolute R-band magnitude of fallback disk outflow emission from blue supergiants~(BSG).  
Each curve corresponds to different outflow launching radius, $f_r = 10, 30,$ and $100$, 
which corresponds to a different outflow speed in our model, $\bar v_{\rm out} \sim 0.32, 0.18, 0.10 c$, respectively (Eqs. \ref{eq:r_0} and \ref{eq:var_v_out}).
Other model parameters are the same as in Table \ref{table:para}.  
}
\label{fig:Rab_fr}
\end{figure}

Fig. \ref{fig:rph_Tph_Lbol} shows the time evolution of the diffusion radius, $r_{\rm dif}$ (top), observed temperature, $T_{\rm obs}$ (middle), and bolometric luminosity $L_{\rm bol}$ (bottom) during $t_{\rm acc} < t < 2t_{\rm p}$. 
The solid red and dotted-dash blue lines show our fiducial model of WR and BSG progenitor (Table \ref{table:para}).  
Up to $t \lesssim t_{\rm p}$ (Eq. \ref{eq:t_tr}), $r_{\rm dif}$, $T_{\rm obs}$, and $L_{\rm bol}$ approximately evolves 
as Eqs. (\ref{eq:rph_homo}), (\ref{eq:Tph_homo}), and (\ref{eq:Lbol_homo}), respectively.
Fig. \ref{fig:Rab} shows the absolute AB magnitude of the fallback disk outflow.  
The thick and thin lines show NUV ($\lambda_{\rm eff} = 2316$ \AA) and R band light curves, respectively.  
The rise time of the UV-optical counterpart is $\sim \rm  a \ few \ days$, which is roughly given by 
the time over which radiation can diffuse across the entire homologous shell in an expansion time ($t_{\rm p}$; Eq. \ref{eq:t_tr}).
The peak bolometric luminosity is $\sim 10^{42-43} \ \rm erg \ s^{-1}$ (Eq. \ref{eq:Lbol_tr}), and the observed temperature is $\sim 10^4 \ \rm K$ (Eq. \ref{eq:Tph_tr}).
The peak absolute magnitude can range from $\sim -16$ to $\sim -18$. 
These events can thus be observed as rapidly-evolving luminous blue transients. 
Fig. \ref{fig:Rab_fr} shows the dependence of the optical (R-band) light curve on the outflow launching radius, 
which in our model sets the speed of the outflow (Eqs. \ref{eq:r_0}-\ref{eq:var_v_out}).  
The rise is faster and the flux is larger for a smaller launching radius, i.e., a higher speed outflow.

In the above calculations, we have assumed that the fallback disk outflows expand in a ``clean" circumstellar medium. 
However, massive stars show mass loss throughout their lives.
The interaction with matter ejected prior to core collapse can alter the dynamics of the outflow and the resulting emission characteristics. 
In particular, if the progenitor experiences an intense mass loss of $\dot M_{\rm w} \gtrsim 0.01 \ M_\odot \ \rm yr^{-1}$ a few years before the collapse, 
a comparable mass to that of the disk outflow ($\gtrsim 0.1 \ \rm M_\odot$) is distributed within the emission radius ($\lesssim 10^{15-16} \ \rm cm$) 
given the typical wind velocity of WRs and BSGs, $v_{\rm w} \sim 10^{3} \ \rm km \ s^{-1}$~\citep[e.g.,][]{Crowther_2001}, 
and the fallback disk outflow will be hidden by the previously ejected matter. 
On the other hand, for an observed typical mass loss rate of WRs and BSGs, $\dot M_{\rm w} \sim 10^{-5} \ \rm yr^{-1}$, the effects of the previously ejected matter will not be significant.

\section{Discussion}\label{sec:dis}
Using a simple analytic model, we have calculated the fallback disk outflow emission from the formation of BHs in otherwise failed supernova explosions. 
Fallback disks power outflows whose emission can be observed as a rapidly-evolving ($\sim \rm a \ few \ days$) luminous ($\sim 10^{42-43} \ \rm erg \ s^{-1}$) blue ($T \sim 10^4 \ \rm K$) transient. 
This outflow can be observed only when it is not enshrouded by a quasi-spherical explosion. 
This likely requires compact progenitors, like WRs and BSGs, which have tightly bound envelopes so that the neutrino radiation of the proto-NS binding energy would not lead to significant mass ejection. 
Our simplified treatments of e.g., the fallback disk formation, the (thermo-)dynamics of the outflow, and the transfer of the cooling radiation, need to be followed up by more detailed numerical studies. 


In the last decade, a growing number of fast transients have been detected by high-cadence surveys 
e.g., Pan-STARRS~\citep{Hodapp_et_al_2004}, PTF~\citep{Law_et_al_2009}, ASAS-SN~\citep{Shappee_et_al_2014}, and LOSS~\citep{Filippenko_et_al_2001}.
For example, the Pan-STARRS1 Medium Deep Survey (PS1-MDS) recently reported a new class of optical transient, which have peak bolometric luminosities of $\sim 10^{42-44} \ \rm erg \ s^{-1}$ 
and shorter decline timescales ($< 15 \ \rm days$) than any type of conventional SNe~\citep{Drout_et_al_2014}.
The spectra can be fitted by blue continua with a temperature of $\sim (1\mbox{-3}) \times 10^{4} \ \rm K$, 
and the lack of UV line blanketing in the spectra imply that the main energy source is not the radioactive decay of ${}^{56}{\rm Ni}$. 
These emission characteristics are broadly consistent with the fallback disk outflow emission from compact progenitors 
with $R_* \sim 10^{12} \ \rm cm$ and $M_{\rm d} \sim$ a few $M_\odot$ (see Eqs. \ref{eq:t_tr}, \ref{eq:Tph_tr}, and \ref{eq:Lbol_tr}).
The host galaxies of the PS1-MDS transients are star-forming galaxies, which is also consistent with our scenario. 
The observed rate of the PS1-MDS transients is $4\mbox{-}7 \ \%$ of core-collapse SN rate at $z = 0.2$  
\citep[cf., the rate of type Ibc SN is $26 \ \%$ of core-collapse SN rate:][]{Smith_et_al_2011}.
In our scenario, this would indicate that fallback disk formation is relatively common in the collapse of WRs and BSGs, 
as theoretically expected~\citep{Woosley_Heger_2012,Perna_et_al_2014}.
We caution, however, that there are likely multiple classes of fast blue transients. 
For example, similar fast transients also have been detected by PTF~\citep[PTF 09uj;][]{Ofek_et_al_2010}. 
Narrow line features observed in PTF 09uj suggest a dense circumstellar envelope surrounding the photon-emitting shell, which is inconsistent with our scenario.  
Shock breakout from an extended envelope and/or wind is a more plausible explanation.  

In our model, the optical-UV emission is powered by thermal energy generated in the accretion disk close to the central black hole. 
The composition of the outflow in turn depends on the temperature reached as the disk circularizes at small radii (Eq. \ref{eq:T_0}).   
For typical parameters, this is $T_0 \sim 10^{8-9} \ \rm K$ indicating that much of the ejecta will be processed to C, O, Ne, and Mg, but not heavier elements.  
Hydrogen and helium may be present depending on the initial composition of the star and the efficiency with which the disk-powered outflow entrains infalling material.   
Our default calculations (Figs. 2 and 3) are for fallback disks that circularize at $r \sim 10 \times GM_{\rm BH}/c^2$ and produce outflow velocities of $\sim 100,000$ km/s, 
which would lead to very Doppler-broadened lines.  
Somewhat slower outflows would produce somewhat slower, fainter transients (Fig. \ref{fig:Rab_fr}).

In order to distinguish our scenario from alternative models, multi-wavelength observations are crucial. 
The most promising one is follow-up observations in the radio.    
The kinetic energy in fallback disk outflows from compact progenitors can be as large as $\approx 0.5 f_{\dot M} M_{\rm d} v_{\rm out, min}^2 \sim 10^{52} \ \rm erg$, 
which dissipates predominately at the forward shock between the outflow and the interstellar medium~(ISM) 
in a deceleration time $\sim (f_{\dot M} M_{\rm d}/m_{\rm p} n_{\rm ISM}v_{\rm out, min}^3)^{1/3} \sim$ a few years. 
Here, $n_{\rm ISM} \sim 1 \ \rm cm^{-3}$ is the number density in the ISM. 
At the forward shock, electrons are accelerated up to a sufficiently high energy and produce broad band synchrotron emission.  
Based on the standard model~\citep{Chevalier_1998, Nakar_Piran_2011}, the peak flux can be $\sim 0.1 \ \rm mJy$ at $\sim \rm GHz$ from $z \sim 0.2$, 
which is detectable by current radio observatories. 
Such a bright radio counterpart cannot be expected in alternative scenarios, e.g., shock breakout in an extended stellar envelope.  
In the above estimate, we assumed that the power law index of non-thermal electrons is $p = 3$, 
and that the acceleration and magnetic-field amplification efficiency are $\epsilon_{\rm e} = \epsilon_{\rm B} = 0.1$, respectively. 

Higher energy counterparts could also help discriminate among the different scenarios.
Simultaneous detections of the UV and optical counterparts predicted in Fig. \ref{fig:Rab} can give more stringent constraints on the model parameters.    
Also, in the early expansion phase of the outflow, the typical energy of the cooling radiation is in the soft X-ray band, $\sim 0.1-10 \ \rm keV$.
The luminosity is typically $\gtrsim 10^{43}  \ \rm erg \ s^{-1}$ lasting for $\sim 100 \ \rm s$. 
For $z \sim 0.2$, the anticipated flux is $\sim 10^{-13} \ \rm erg \  s^{-1} \ cm^{-2}$, which could be detectable by eROSITA~\citep{Merloni_et_al_2012}. 


\section*{Acknowledgments}
We thank Rodrigo Fernandez, Weikang Zheng, Nobuya Nishimura, and Shota Kisaka for useful discussions. 
We also thank Daniel Kasen, Christopher S. Kochanek, Anthony Piro, Christian D. Ott, and Brian Metzger for valuable comments.
KK is supported by NASA through Einstein Postdoctoral Fellowship grant number PF4-150123 awarded by the Chandra X-ray Center, 
which is operated by the Smithsonian Astrophysical Observatory for NASA under contract NAS8-03060. 
EQ was supported in part by NSF grant AST-1205732, a Simons Investigator award from the Simons Foundation and the David and Lucile Packard Foundation. 


\label{lastpage}

\end{document}